\documentclass[prl,twocolumn,floatfix,superscriptaddress,showpacs]{revtex4}

\usepackage{graphicx}

\begin{document}

\title{Josephson effect in superfluid atomic Fermi-gases}

\date{\today}

\author{Gh.-S.\ Paraoanu}
\altaffiliation[also at the ]{Department
        of Theoretical Physics, National Institute for Physics and Nuclear
        Engineering, PO BOX MG-6, R-76900, Bucharest, Romania.}
\affiliation{Department of Physics, University of
Jyv\"askyl\"a, P.O.Box 35, FIN-40351 Jyv\"askyl\"a, Finland}
\affiliation{Department of Physics, Loomis Laboratory, 1110 W.\ Green
Street,
University of Illinois at Urbana-Champaign, Urbana IL61801, USA}

\author{M.\ Rodriguez}
\affiliation{Laboratory of Computational Engineering, P.O.Box 9400, 
FIN-02015
Helsinki University of Technology, Finland}

\author{P.\ T\"orm\"a}
\affiliation{Department of Physics, University of Jyv\"askyl\"a, P.O.Box 
35,
FIN-40351 Jyv\"askyl\"a, Finland}

\begin{abstract}

We consider an analog of the internal Josephson effect in superfluid 
atomic Fermi-gases.
Four different hyperfine states of the atoms are assumed to be
trapped and to form two superfluids {\it via} the BCS-type of pairing.  
We show that Josephson oscillations can be realized by coupling
the superfluids with two laser fields. Choosing the laser detunings
in a suitable way leads to  
an asymmetric below-gap tunneling effect
for which there exists no analogue in the 
context of solid-state superconductivity. 

\end{abstract}


\maketitle

  Cooling of trapped gases of Fermionic atoms well below the Fermi
  temperature \cite{Debbie,Hulet,Salomon,Thomas,KetterleF}  makes it reasonable to
  anticipate the achievement of the predicted BCS-transition
  \cite{Stoof,Holland,Zoller,Pethick}.  The existence of a gap in the excitation
  spectrum of the superfluid Fermi-gas will be the first issue to
  address, and several methods for detecting it have already been
  proposed \cite{alli,ours}.  Trapped atomic Fermi-gases will allow to
  study and test fermion-fermion pairing 
  theories in a tunable, controlled manner. For example
  the classic problem of the BCS-BEC crossover when the interparticle
  attraction varies \cite{bcsbec} could be studied using the
  possibility of tuning the interatomic scattering length using 
Feshbach resonances \cite{Holland,feshbach}. 
Besides the standard superfluid phenomenology, 
gases of Cooper-paired atoms are expected to have properties which are
  specific to atomic gases only and not present, or not easily realizable,
  in metallic superconductors or Helium. For instance, the trapping potential
has a major effect on the characteristic lengths of the 
superfluid Fermi-gas \cite{vortex}. 

In this paper we propose a way to investigate the
Josephson 
effect in trapped superfluids of Fermionic atoms. We find a phenomenon 
  that is unique to atomic
  Fermi-gases, namely an asymmetry in the Josephson currents 
corresponding to the "up" and "down" spin states. 
We assume that Fermionic atoms in four different hyperfine states (we label
them $|g\rangle$, $|g'\rangle$, $|e\rangle$ and $|e'\rangle$) are trapped
simultaneously in an optical trap --- recently  all-optical 
trapping and cooling below the
degeneracy point of the two lowest hyperfine states of $^6$Li has been 
demonstrated \cite{Thomas}.
The s-wave scattering lengths are assumed to be large 
and negative between atoms 
in states
$|g\rangle$ and $|g'\rangle$, as well as between those in $|e\rangle$ 
and $|e'\rangle$,
and the chemical potentials 
$\mu_g \simeq \mu_{g'}$ and $\mu_e \simeq \mu_{e'}$.
For all other combinations of two atoms in different states 
the scattering length
is assumed to be small and/or the chemical potentials unequal. This leads to
the existence of two superfluids, one consisting of Cooper pairs of atoms
in the states $|g\rangle$ and $|g'\rangle$, and the other of 
$|e\rangle$--$|e'\rangle$
pairs. The configuration is experimentally challenging, but
in principle possible by the choice of right atoms and hyperfine states,
adjusting the number of atoms, and tuning the scattering lengths in 
magnetic fields by using Feshbach resonances \cite{Holland,feshbach}. 

The two superfluids are coupled 
by driving laser-induced transitions between the states $|g\rangle$ and 
$|e\rangle$
with the laser Rabi frequency $\Omega$ and detuning $\delta$, and between
the states $|g'\rangle$ and $|e'\rangle$ with the Rabi frequency $\Omega'$ 
and detuning $\delta'$. For a Raman-process these are effective
quantities. If several 
lasers are used then in order to be able to see the Josephson
oscillations they should maintain their phase coherence for a
time much longer than the inverse of the detunings. 
In the case of metals, 
the two superconductors are spatially separated and connected by a 
tunneling junction. In our scheme, 
the superfluids share 
the same spatial region 
and are connected by the laser-coupling of the atoms' internal states; 
this resembles the internal Josephson effect 
in atomic Bose-Einstein condensates \cite{internal} 
or in superfluid $^3$He-A \cite{leggett}.

For metallic 
superconductors the a.c. Josephson current is driven by 
applying a voltage over the junction -- here the role of the voltage is 
played by the laser detunings. The difference is that the detunings 
can be different for the 
two states forming the pair; in the metallic superconductor analogy 
this would mean having a different voltage for the 
spin-up and spin-down electrons, a situation which has 
not been investigated in the context 
of metallic superconductors. There is an interesting connection
to recent experiments on superconductor-ferromagnet
proximity effects, where the chemical potentials of the spin
up and down electrons are slightly different in the ferromagnet 
due to the exchange interaction \cite{ferro}.


We consider a system described by the standard BCS-theory. The laser
interaction is assumed to be a small perturbation and its effect is calculated
using linear response theory. The observable of interest is the change in
the number of particles in one of the states, say $|e\rangle$ or $|e'\rangle$.

\begin{figure}
\includegraphics[scale=0.65]{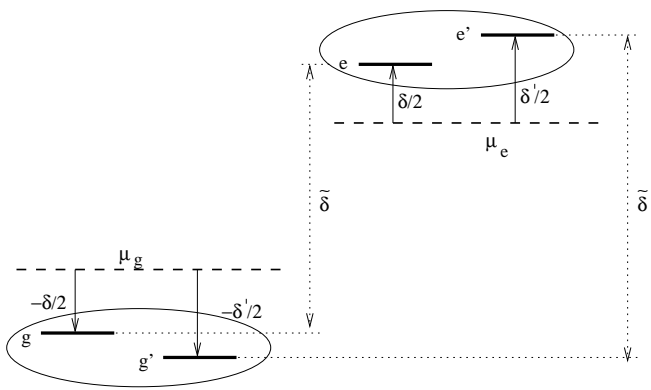}
\caption{The effect of the detunings $\delta$ and $\delta '$ of the two 
lasers. The initial chemical potentials (in the absence of laser 
couplings) were $\mu_{g}=\mu_{g'}$ and $\mu_{e}=\mu_{e'}$. The states
$(gg')$ and $(ee')$ are Cooper-paired.}
\label{fj_fig1}
\end{figure}

In the rotating wave approximation the interaction of the laser light
with the matter fields can be described by a time-independent 
Hamiltonian in which the detunings $\delta$ and $\delta'$ play the role of an
externally imposed difference in the chemical potential
of the two states. The total Hamiltonian becomes then $\hat{H} = 
\hat{H_0} +
\hat{H}_{T}$, where
\begin{widetext}
\begin{eqnarray}
 \hat{H_0} &=& \hat{H}_{BCS}
+\left(\mu_{e} + \frac{\delta}{2}\right)\int 
d\vec{r}
\hat{\psi}_e^{\dagger}(\vec{r})\hat{\psi}_e
(\vec{r})
+ \left(\mu_{g} - \frac{\delta}{2}\right)\int d\vec{r}
\hat{\psi}_g^{\dagger}(\vec{r})\hat{\psi}_g
(\vec{r})   \nonumber \\
&& +\left(\mu_{e} + \frac{\delta'}{2}\right)\int 
d\vec{r}
\hat{\psi}_{e'}^{\dagger}(\vec{r})\hat{\psi}_{e'}
(\vec{r})
+ \left(\mu_{g} - \frac{\delta'}{2}\right)\int d\vec{r}
\hat{\psi}_{g'}^{\dagger}(\vec{r})\hat{\psi}_{g'}
(\vec{r}) \nonumber. 
\end{eqnarray}
\end{widetext}
Here $\mu_{e}$ and $\mu_{g}$ are the chemical potentials of the Fermi 
gases before 
the laser was turned on, $\mu_g =\mu_{g'}$ and $\mu_e = \mu_{e'}$ 
in order to allow standard BCS pairing.
The Hamiltonian $\hat{H}_{BCS}$ is the BCS-approximation 
of the matter Hamiltonian with 
the chemical potential included \cite{boogie}. 
Figure {\ref{fj_fig1}} presents a schematic view of what happens
in this case: the laser detunings shift the chemical potentials of 
the four hyperfine states.

The transfer Hamiltonian is given by 
\begin{eqnarray}
\hat{H}_{T} &=& \int d\vec{r}\Omega (\vec{r})\hat{\psi}_e
^{\dagger}(\vec{r})\hat{\psi}_g(\vec{r}) + \Omega^{*}
(\vec{r})\hat{\psi}_g^{\dagger}(\vec{r})
\hat{\psi}_e(\vec{r}) \nonumber \\
&+& \int d\vec{r}\Omega' (\vec{r})\hat{\psi}_{e'}
^{\dagger}(\vec{r})\hat{\psi}_{g'}(\vec{r}) + \Omega'^{*}
(\vec{r})\hat{\psi}_{g'}^{\dagger}(\vec{r})
\hat{\psi}_{e'}(\vec{r}),
\end{eqnarray}
with $\Omega (\vec{r})$ and $\Omega' (\vec{r})$ characterizing 
the local strength of the 
matter-field interaction.  

The main observable of interest, the 
rate of transferred atoms from, say, state $|g\rangle$ to 
state $|e\rangle$, is defined by
\begin{equation}
I_e = \frac{\partial}{\partial t}\int d\vec{r}\langle\Psi 
(t)|\hat{\psi}_e^{\dagger}(\vec{r})\hat{\psi}_e(\vec{r})
|\Psi (t)\rangle 
\end{equation}
(the definition for $I_{e'}$ is similar)
and can be further evaluated with the help of the Schr\"odinger equation
$i\hbar\frac{\partial}{\partial t}|\Psi (t)\rangle = \hat{H} |\Psi (t)
\rangle$ as
\begin{equation}
I_e = i \int d\vec{r}\langle\Psi (t)|
\Omega^{*}(\vec{r})\hat{\psi}_g^{\dagger}(\vec{r})
\hat{\psi}_e(\vec{r}) - \Omega (\vec{r})
\hat{\psi}_e^{\dagger}(\vec{r})
\hat{\psi}_g(\vec{r})|\Psi (t)\rangle.
\end{equation}
In the following we call $I_e$ the current in analogy to
metallic superconductors where the flux of electrons out of the
superconductor constitutes the electrical current.

We introduce an interaction representation with respect to 
$\hat{H_0}$ and use linear response theory with 
respect to $\hat{H}_{T}$. 
Validity of the linear response theory requires that
the laser intensity is small and the transfer of atoms can be
treated as a perturbation. 

We split the result for the current $I_e$ into a part which corresponds to
the Josephson current $I_{eJ}$ and to the part 
which describes normal single-particle current
$I_{eS}$, $I_e = I_{eJ} + I_{eS}$. The single-particle current can be 
evaluated at finite temperature using the standard techniques of 
superconducting Green's functions; we present however only the 
result for $T=0$ and positive detunings: 
\begin{eqnarray}
I_{eS} &=& -2\pi\sum_{n,m}\left|\int d\vec{r}\Omega 
(\vec{r})v^e_{n}(\vec{r})u^g_{m}(\vec{r})\right|^{2}\delta (\epsilon^e_{n} + 
\epsilon^g_{m} - \tilde{\delta} ) . \nonumber
\end{eqnarray}
Here the triplet $(u_{n}, v_{n}); \epsilon_{n}$ is a solution of the 
(nonuniform) Bogoliubov-de Gennes equations for superconductors 
\cite{boogie} and $\tilde{\delta} = \mu_e - \mu_g + \delta$. This is 
the 
standard Fermi Golden rule result and very similar to the ones 
obtained in \cite{ours}. The current $I_{eS}$ is zero when
$\delta < 2 \Delta$ since pair breaking is required for 
single particle excitations. 
Next we concentrate on the Josephson current -- this is non-zero 
also for detunings $\delta$ that are smaller than twice the gap energy. 

The Josephson current becomes 
\begin{widetext}
\begin{equation} 
I_{eJ} = -2 {\rm Im}
\left[ e^{-i(\tilde{\delta} + \tilde{\delta}')t} \sum_{n,m}\int 
d\vec{r}d\vec{r}'\Omega^{*}(\vec{r}) \Omega'^{*} (\vec{r}') 
u_{n}^{g*}(\vec{r})u^e_{m}(\vec{r}')v_{m}^{e*} 
(\vec{r})v^g_{n}(\vec{r}')
\left(\frac{1}{\tilde{\delta}' + 
\epsilon^g_{n} +
\epsilon^e_{m} + i\eta} -\frac{1}{\tilde{\delta}' - \epsilon^g_{n} -
\epsilon^e_{m} + i\eta} \right)  \right].\label{doibre} 
\end{equation}
\end{widetext}
The current $I_{e'J}$ is the same only that $\tilde{\delta}$ and $\tilde{\delta'}$ are
interchanged. Note that the oscillating term is proportional to both of
the detunings whereas the rest of the expression is proportional 
only to $\tilde{\delta'}$. For
the choice of a homogeneous system (large trap, local density
approximation) and a constant laser profile the expression simplifies
into \begin{eqnarray} I_{eJ} &=& I_0(\tilde{\delta}') \sin 
[(\tilde{\delta} + \tilde{\delta}')t]
\\ I_{e'J} &=& I_0(\tilde{\delta}) \sin [(\tilde{\delta} + 
\tilde{\delta}')t] . \end{eqnarray}
Both partners of the pair thus oscillate in phase, with the same
frequency $\tilde{\delta} + \tilde{\delta}'$. But the amplitudes are 
different whenever
the detunings $\tilde{\delta}$ and $\tilde{\delta}'$ differ. This means 
that more atoms
are transferred, say, in the $|g\rangle - |e\rangle$ oscillation than in
the $|g'\rangle - |e'\rangle$ one. 

\begin{figure}
\includegraphics[scale=0.37]{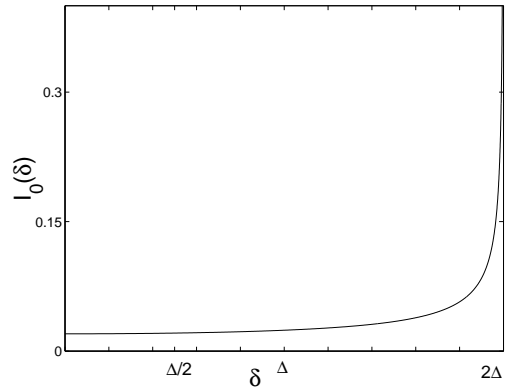}
\caption{The current $I_{0}$ (arbitrary units) as a function of detuning
$\delta$, given in units of $\Delta$.}
\label{fj_fig2}
\end{figure}

A simple expression for $I_0(\delta)$
can be derived when we assume identical superfluids, that is $\Delta =
\Delta'$ and $\mu_{g} = \mu_{e} \equiv \mu$: 
\begin{eqnarray} 
I_0(\delta)
= \frac{\sqrt{2m^3}V}{\pi^{2}}\Delta^2 \Omega^2 \int_{-\mu}^\infty  
\frac{d\xi\sqrt{\mu +
\xi}}{\sqrt{\xi^2 + \Delta^2} (4\xi^2 + 4 \Delta^2 - 
\delta^2)}, \nonumber
\end{eqnarray} 
where $V$ is the volume of the sample and the variable $\xi$ is the 
continuous version of $\xi_{k}=\frac{k^{2}}{2m}-\mu$. Since $\Delta \ll 
\mu$, the result can also be written as
\begin{eqnarray}
I_0(\delta)
= \frac{\sqrt{2m^3\mu}V}{\pi^{2}}\Delta^2 \Omega^2 
\int_{-\infty}^\infty
\frac{d\xi}{\sqrt{\xi^2 + \Delta^2} (4\xi^2 + 4 \Delta^2 -   
\delta^2)}.\nonumber
\end{eqnarray}
A plot of the intensity $I_{0}$ as a function of the detuning $\delta$ 
is shown in Fig. {\ref{fj_fig2}}.
This result shows a divergence at $\delta = 2\Delta$, which 
reflects the divergence of the density of states for the two superconductors 
at the gap. For quite a large range of detunings, 
the amplitude $I_{0}$ of the Josephson current is approximately 
constant --- thus no asymmetry effect will be visible. The asymmetry is
most pronounced when the timescale of the oscillation, that is, $1/(\delta
+ \delta')$ is close to $1/(2\Delta)$. Note that $1/(2\Delta)$ 
can also be understood as the Cooper pair correlation time based
on the uncertainty principle.

According to the conventional intuitive picture of the Josephson effect,
the particles forming a Cooper-pair tunnel ``together'' through the junction.
Therefore our result seems counterintuitive at first glance.
The physics becomes, however, more transparent by a closer look at the equation
(\ref{doibre}). For simplicity, we consider here the transfer process to one direction only,
from the superfluid $(gg')$ to $(ee')$, which corresponds to the first denominator
in (\ref{doibre}). In the initial state $|g\rangle$ is paired with $|g'\rangle$, in the final
state $|e\rangle$ with $|e'\rangle$. The process has, however, an intermediate state as indicated
by the second-order form of the observable, and the intermediate states corresponding to
the observables $I_{eJ}$ and $I_{e'J}$ are different: For $I_{eJ}$,  
$|g'\rangle$ has been transferred into $|e'\rangle$. Therefore its pairing partner
$|g\rangle$ is left as an excitation in the superfluid $(gg')$ with the energy 
$\epsilon_{n}^{g}$ and $|e'\rangle$ becomes an excitation in the superfluid $(ee')$ with the energy
$\epsilon_{m}^{e'}=\epsilon_{m}^{e}$. 
In contrast, for $I_{e'J}$, the atom in $|g'\rangle$ remains as a quasiparticle of the energy $\epsilon_{n}^{g'}
= \epsilon_{n}^{g}$ in the superfluid $(gg')$ and
$|e\rangle$ becomes an excitation in the superfluid $(ee')$. 
For $I_{eJ}$, the initial 
energy of the Cooper pair was $\left(\mu_{g} - \frac{\delta}{2}\right) + 
\left(\mu_{g}-\frac{\delta '}{2}\right)$ and the energy of the 
intermediate state is 
$\left[\left(\mu_{g}-\frac{\delta }{2}\right) + 
\epsilon_{n}^{g}\right] + 
\left[\left(\mu_{e} + \frac{\delta '}{2}\right) 
+\epsilon_{m}^{e}\right]$ (for explanation, see Fig.{\ref{fj_fig1}}). 
The relative energy of the intermadiate state with respect to the initial state is 
$\epsilon_{n}^{g} +
\epsilon_{m}^{e} +\tilde{\delta}'$, which is precisely the first denominator in 
(\ref{doibre}). For $I_{e'J}$, the initial 
energy of the pair is again $\left(\mu_{g} - \frac{\delta}{2}\right) +
\left(\mu_{g}-\frac{\delta '}{2}\right)$, but the intermediate state
has an energy $\left[\left(\mu_{g}-\frac{\delta '}{2}\right) +
\epsilon_{n}^{g}\right] +
\left[\left(\mu_{e} + \frac{\delta }{2}\right)
+\epsilon_{m}^{e}\right]$, 
or a relative energy
$\epsilon_{n}^{g} +
\epsilon_{m}^{e} +\tilde{\delta}$. In summary, the intermediate states
of the transfer processes for ``spin up'' and ``spin down'' atoms 
have different energies and this results in different amplitudes for $I_{eJ}$ and $I_{e'J}$.  

The asymmetry in the currents implies the
existence of excitations in the superfluids. 
We have analyzed the many-body wavefunction of the system in the
Schr\"odinger picture and indeed it contains excitations corresponding
to the asymmetry. Specifically, the analysis confirms that 
the so called Fermi surface polarization
$(\langle N_e \rangle - \langle N_{e'} \rangle)/(\langle N_e \rangle + \langle N_{e'} \rangle)
\simeq (\langle N_e \rangle - \langle N_{e'} \rangle)/
(\langle N_e \rangle_0 + \langle N_{e'} \rangle_0)$ is non-zero
and oscillates as $f(\delta, \delta') \cos((\delta+\delta')t)$ where
$f(\delta, \delta) = 0$. We also found that {\it time-independent} perturbation
theory is not sufficient to reveal the asymmetry in the amplitudes: the
simple treatment of \cite{timeind} applied to our system results in symmetric
currents because the ansatz used does not allow any excitations.
This and the fact that the oscillation is most pronounced for timescales
of the order of the Cooper-pair correlation time indicates that the effect
is related to the dynamics of the superfluid state. 

To observe the Josephson effect one should be able to measure the number
of particles in two of the states, e.g.\ $|e \rangle$ and $|e' \rangle$,
at different stages of the oscillations, either destructively or
non-destructively. The scale of the gap energy is for typical systems
1-100kHz, which means that the highest time resolution needed should be
just somewhat above $10 \mu s$. Measuring the number of particles
accurately is the more challenging part of the observation.  
In \cite{ours} we considered laser probing of the superfluid
Fermi-gas, where the laser was creating excitations in the BCS state.
The number of particles transferred was directly reflected in the
absorption of the light. Here one can use similar techniques to detect
the Josephson oscillations in a simple way.

In summary, we propose a method to realize Josephson oscillations in
superfluid atomic Fermi-gases. The coupling between two superfluids is
provided by laser light, and the laser detuning plays the same role as
voltage over metallic superconductor junctions. Detunings that affect
the two atomic internal states involved in pairing can be chosen to be
different -- this would correspond to different voltage for spin-up and
spin-down electrons. This leads to asymmetry in the oscillation
amplitudes of the two states. The asymmetry is pronounced when the
time-scale of the oscillation is the same order of magnitude as the
Cooper-pair correlation time. 
This is an effect unique to atomic Fermi-gases in the superfluid state.

\begin{acknowledgments} 
We thank the Academy of Finland for support
(projects 42588, 48845, 47140 and 44897). Gh.-S. P. also acknowledges
the grant NSF DMR 99-86199. We wish to thank Prof. G. Baym and Prof. A.
J. Leggett for useful discussions.
\end{acknowledgments}


\begin{thebibliography}{99}

\bibitem{Debbie}
B.\ DeMarco and D.S.\ Jin, Science \textbf{285}, 1703 (1999); 
M.J.\ Holland, B.\ DeMarco, and D.S.\ Jin, 
Phys.\ Rev.\ A \textbf{61}, 053610 (2000); B.\ DeMarco, S.B.\ Papp,
and D.S.\ Jin, Phys.\ Rev.\ Lett.\ {\bf 86}, 5409 (2001).

\bibitem{Hulet}
A.G. Truscott, K. E. Strecker, W. I. McAlexander,
G. P. Partridge, and R. G. Hulet, Science \textbf{291}, 2570 (2001).

\bibitem{Salomon}
M.O.\ Mewes, G.\ Ferrari, F.\ Schreck, A.\ Sinatra, and C.\ Salomon, Phys.\ 
Rev.\ A \textbf{61}, 011403 (R) (2000); F.\ Schreck, L.\ Khaykovich, K.L.\ 
Corwin, G.\ Ferrari, T.\ Bourdel, J.\ Cubizolles, and C.\ Salomon,
Phys.\ Rev.\ Lett.\ {\bf 87}, 080403 (2001).

\bibitem{Thomas}
K.M.\ O'Hara, M.E.\ Gehm, S.R.\ Granade, S.\ Bali, and J.E.\ Thomas, 
Phys.\ Rev.\ Lett.\ \textbf{85}, 2092 (2000);
S.R.\ Granade, M.E.\ Gehm, K.M.\ O'Hara, and J.E.\ Thomas,
Phys.\ Rev.\ Lett.\ {\bf 88},120405 (2002). 

\bibitem{KetterleF}
Z.\ Hadzibabic, C.A.\ Stan, K.\ Dieckmann, S.\ Gupta, M.W.\ Zwierlein,
A.\ G\"orlitz, and W.\ Ketterle, Phys.\ Rev.\ Lett.\ \textbf{88}, 160401 (2002).

\bibitem{Stoof}
H.T.C.\ Stoof,  M.\ Houbiers, C.A.\ Sackett, and R.G.\ Hulet, 
Phys.\ Rev.\ Lett.\ \textbf{76}, 10 (1996); 
M.\ Houbiers, R.\ Ferwerda, H.T.C.\ Stoof, W.I.\  McAlexander, C.A.\ 
Sackett, and R.G.\ Hulet, Phys.\ Rev.\ A \textbf{56}, 4864, (1997); 
R.\ Combescot, Phys.\ Rev.\
Lett.\ {\bf 83}, 3766 (1999). 

\bibitem{Holland}
M.\ Holland, S.J.J.M.F.\ Kokkelmans, M.L.\ Chiofalo, 
and R.\ Walser, Phys.\ Rev.\ Lett.\ {\bf 87}, 120406 (2001);
J.N.\ Milstein, S.J.J.M.F.\ Kokkelmans, and M.J.\ Holland, cond-mat/0204334
(2002). 

\bibitem{Zoller}
W.\ Hofstetter, J.I.\ Cirac, P.\ Zoller, E.\ Demler, and M.D.\ Lukin,
cond-mat/0204237 (2002).

\bibitem{Pethick}
H.\ Heiselberg, C.J.\ Pethick, H.\ Smith, and L.\ Viverit, Phys.\ Rev.\
Lett.\ \textbf{85}, 2418 (2000). 

\bibitem{alli}
W.\ Zhang, C.A.\ Sackett, and R.G.\ Hulet, Phys.\ Rev.\ A \textbf{60}, 504 (1999); 
J.\ Ruostekoski, Phys.\ Rev.\ A \textbf{60}, R1775 (1999); 
F.\ Weig and W.\ Zwerger, Europhys.\ Lett.\ \textbf{49}, 282 (2000); M.A.\ Baranov
 and D.S.\ Petrov, 
Phys.\ Rev.\ A \textbf{62}, 041601(R) (2000); M.\ Farine, P.\ Schuck,
and X.\ Vi\~nas,
 Phys.\ Rev.\ A \textbf{62}, 013608 (2000); G.M.\ Bruun and C.W.\
 Clark, J.\ Phys.\ B \textbf{33}, 3953 (2000).

\bibitem{ours}
P.\ T\"orm\"a and P.\ Zoller, Phys.\ Rev.\ 
Lett.\ \textbf{85}, 487 (2000); G.M.\ Bruun, P.\ T\"orm\"a,
M.\ Rodriguez, and P.\ Zoller,,
Phys.\ Rev.\ A \textbf{64}, 033609 (2001); Gh.-S.\ Paraoanu, M.\ Rodriguez, 
and P. T\"orm\"a, J.\ Phys.\ B: At.\ Mol.\
Opt.\ Phys.\ \textbf{34}, 4763 (2001). 

\bibitem{bcsbec}
See M.\ Randeria and references therein in {\it Bose-Einstein Condensation},
eds.\ A.\ Griffin, D.W.\ Snoke, and S.\ Stringari (Cambridge Un.\ Press,
Cambridge, 1995).

\bibitem{feshbach} E.\ Tiesinga, B.\ J.\ Verhaar, and H.\ T.\ C.\ Stoof, 
Phys. Rev. 
A \textbf{47}, 4114 (1993); S.\ Inouye, M.R.\ Andrews, J. Stenger, H.-J.\ Miesner,
D.M.\ Stamper-Kurn, and W.\ Ketterle, Nature {\bf 392}, 
151 (1998); J.L.\ Roberts, N.R.\ Claussen, S.L.\ Cornish, E.A.\ Donley, E.A.\ Cornell,
and C.E.\ Wieman, Phys.\ Rev.\ Lett.\ \textbf{86}, 4211; 
E. Timmermans, K.\ Furuya, P.W.\ Milonni, and A.K.\ Kerman, Phys. Lett. A {\bf 285}, 228 
(2001).  

\bibitem{vortex}
M.\ Rodriguez, G.-S.\ Paraoanu, and P.\ T\"orm\"a, Phys.\ Rev.\ Lett.\ {\bf 
87}, 100402 (2001).

\bibitem{internal} see
A.\ J.\ Leggett, Rev. Mod. Phys. {\bf 73}, 307 (2001) and references 
therein.

\bibitem{leggett} A.\ J.\ Leggett, Rev. Mod. Phys.{\bf 47}, 331 (1975);
J.\ C.\ Weatley, Rev. Mod. Phys. {\bf 47}, 415 (1975). 

\bibitem{ferro}
T.\ Kontos, M.\ Aprili, J.\ Lesueur, F.\ Genet, B.\ Stephanidis, and R.\ Boursier,
cond-mat/0201104 (2002); F.S.\ Bergeret, A.F.\ Volkov, and K.B.\ Efetov, 
Phys.\ Rev.\ Lett.\ \textbf{86}, 3140 (2001); E.A.\ Demler, G.B.\ Arnold, and M.R.\ Beasley,
Phys.\ Rev.\ B \textbf{55}, 15174 (1997).

\bibitem{boogie}
P.\ de Gennes, {\it Superconductivity of metals and alloys} (Addison- Wesley,
New York, 1966).

\bibitem{timeind}
R.A.\ Ferrell and R.E.\ Prange, Phys.\ Rev.\ Lett.\ \textbf{10}, 479 (1963).

\end{thebibliography}
\end{document}